\documentclass[amsmath,amssymb,twocolumn,superscriptaddress,floatfix]{revtex4-1}
\usepackage{graphicx}
\usepackage{dcolumn}
\usepackage{bm}
\usepackage{hyperref}
\usepackage{multirow}
\usepackage{color}
\usepackage{textgreek}

\begin{document}
\title{Multiple intrinsically identical single photon emitters in the solid-state}

\author{L. J. Rogers}
\author{K. D. Jahnke}
\affiliation{Institute for Quantum Optics and Center for Integrated Quantum Science and Technology (IQ${}^{st}$), University Ulm, D-89081 Germany}
\author{T. Teraji}
\affiliation{National Institute for Materials Science, 1-1 Namiki, Tsukuba, Ibaraki 305-0044, Japan}
\author{L. Marseglia}
\author{C. M\"uller}
\author{B. Naydenov}
\author{H. Schauffert}
\affiliation{Institute for Quantum Optics and Center for Integrated Quantum Science and Technology (IQ${}^{st}$), University Ulm, D-89081 Germany}
\author{C. Kranz}
\affiliation{Institute of Analytical and Bioanalytical Chemistry, University Ulm, D-89081 Germany}
\author{J. Isoya}
\affiliation{Research Center for Knowledge Communities, University of Tsukuba, 1-2 Kasuga, Tsukuba, Ibaraki 305-8550, Japan}
\author{L. P. McGuinness}
\author{F. Jelezko}
\affiliation{Institute for Quantum Optics and Center for Integrated Quantum Science and Technology (IQ$\mathrm{^{st}}$), University Ulm, D-89081 Germany}

\begin{abstract}
Emitters of indistinguishable single photons are crucial for the
growing field of quantum technologies \cite{knill2001scheme,
yuan2008experimental, bernien2013heralded}. To realize scalability and increase
the complexity of quantum optics technologies, multiple independent yet
identical single photon emitters are also required. However typical solid-state
single photon sources are inherently dissimilar, necessitating the use of
electrical feedback \cite{patel2010two-photon, bernien2012two-photon,
sipahigil2012quantum} or optical cavities \cite{santori2002indistinguishable}
to improve spectral overlap between distinct emitters. Here, we demonstrate
bright silicon-vacancy (SiV$^-$) centres in low-strain bulk diamond which
intrinsically show spectral overlap of up to 91\% and near transform-limited
excitation linewidths. Our results have impact upon the application of single
photon sources for quantum optics and cryptography, and the production of next
generation fluorophores for bio-imaging.
\end{abstract}

\date{\today}

\pacs{}

\maketitle



Single, transform-limited photons are an essential resource for many quantum
interference experiments, since indistinguishability between photons allows
path-of-origin information to be erased. This makes possible investigation of
fundamental quantum optics phenomena which have applications in quantum imaging
\cite{strekalov1995observation}, quantum computing \cite{knill2001scheme} and
quantum repeaters \cite{yuan2008experimental}.  Interactions between
indistinguishable photons from multiple emitters can also be used to create
entangled quantum states over macroscopic physical distances
\cite{bernien2013heralded}.  However, to date it has proved difficult to
achieve indistinguishability between distinct single photon sources.

Individual trapped ions in vacuum are a natural source of identical photons,
limited primarily by their independent motion \cite{beugnon2006quantum},
whereas in the solid-state it has been necessary to use optical cavities and/or
electrical tuning in order to interfere photons from distinct emitters
\cite{shields2007semiconductor}. In particular, quantum dots and single
molecules have demonstrated transform-limited linewidths over short timescales,
allowing interference between photons from the same emitter
\cite{santori2002indistinguishable,kiraz2005indistinguishable} and physically
separated emitters \cite{patel2010two-photon}. Colour centres in diamond,
located deep within an ultrapure lattice, show excellent photostability and are
attractive single photon sources. The negative nitrogen vacancy (NV$^-$) centre
is well studied and has been used for early photonics applications
\cite{faraon2012coupling,barclay2011hybrid}, but its broad spectral emission is
ill-suited to single photon technologies \cite{bernien2013heralded}.

\begin{figure*}[thb]
\includegraphics[width=\textwidth]{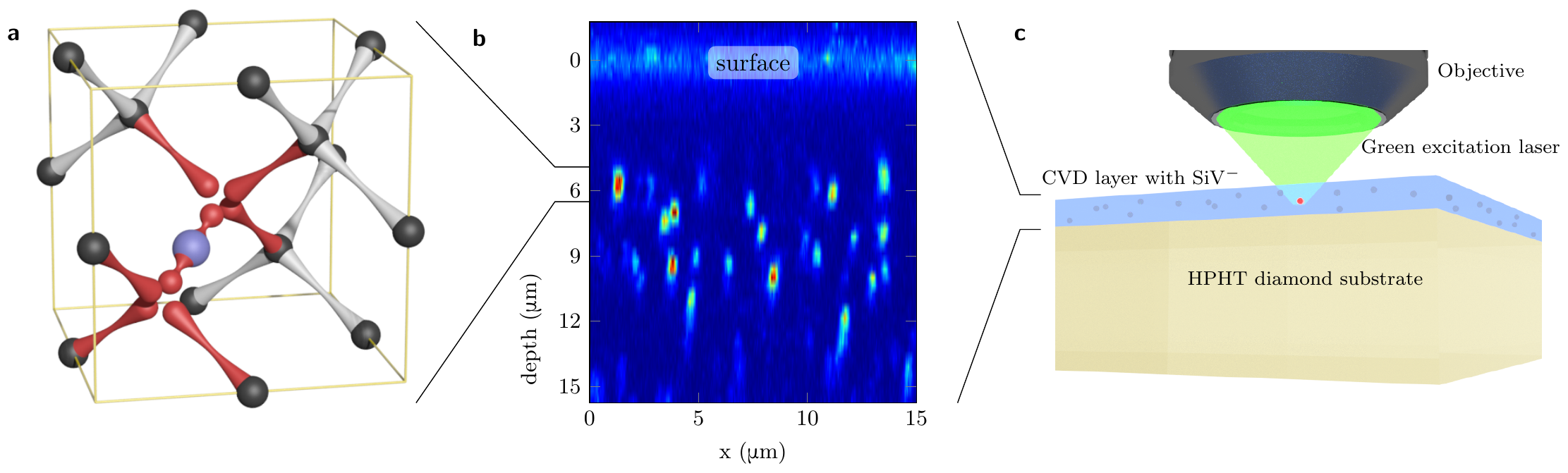}
\caption{Silicon vacancy centers in diamond.
	\textsf{\textbf{a}}. Physical structure of the SiV$^-$ centre in the
	diamond lattice, with a silicon atom lying in between adjacent vacancies.
	\textsf{\textbf{b}}. Fluorescence confocal image showing the depth profile
	of single SiV$^-$ centres beneath the diamond surface (corrected for refractive index).
	\textsf{\textbf{c}}. The SiV$^-$ centers were present in a single crystal
	CVD diamond layer, and had been incorporated during growth due to silicon
	doping of the plasma.  The CVD layer was grown homoepitaxially on low
	strain, high-pressure, high-temperature diamond substrates and investigated
	from the top.
}
\label{fig:fluorescence}
\end{figure*}

The negative silicon-vacancy (SiV$^-$) centre has promising spectral properties
\cite{collins1990spectroscopic, goss1996twelve-line, neu2013low-temperature}.
It has a strong optical transition with a prominent zero-phonon line (ZPL) at
737\,nm and only a weak phonon sideband \cite{feng1993characteristics,
brown1995site, clark1995silicon, edmonds2008electron, neu2011single}.
Structurally, it is comprised of a silicon atom located between adjacent
vacancies in the diamond lattice \cite{goss1996twelve-line,
turukhin1996picosecond, wang2006single} (\autoref{fig:fluorescence}a).  The
strong ZPL has sparked interest, but large variation in spectral properties
between individual sites \cite{neu2013low-temperature} has limited the value of
SiV$^-$ as a single photon emitter.


Here we obtain highly uniform, narrow linewidths by using low strain,
high-pressure, high-temperature (HPHT) diamond as a substrate of high
crystalline quality. A layer of ultrapure diamond was overgrown on this
$\langle001\rangle$-oriented substrate with chemical vapour deposition (CVD),
allowing precise control of impurity concentrations.  Silicon was introduced to
the plasma and incorporated into the diamond at concentrations below 1\,ppb
(see Methods). Single SiV$^-$ defects formed during growth were observed by
fluorescence confocal microscopy (\autoref{fig:fluorescence}b,c).  Photon
antibunching measurements provided confirmation of single photon emission. 

\begin{figure*}[thb]
\includegraphics[width=\textwidth]{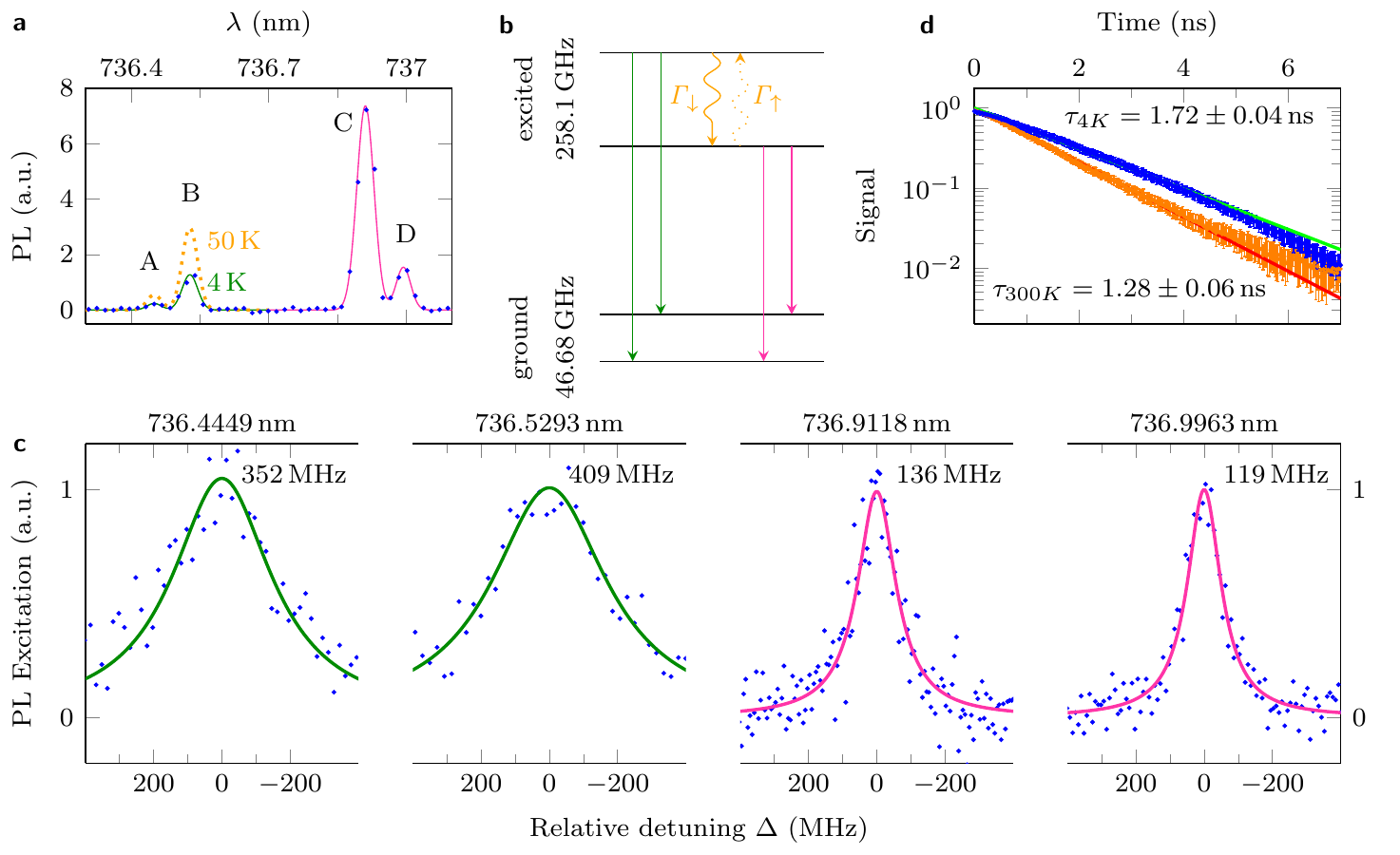}
\caption{Linewidths near the transform-limit.
	\textsf{\textbf{a}}. The SiV$^-$ ZPL has a four-line fine structure
	(linewidth limited here by spectrometer resolution).  The left-hand pair of
	lines are a mirror image of the right-hand pair, but they have much lower
	intensity at 4\,K.  These higher energy lines regain intensity with
	increasing temperature, and the dotted line illustrates the relative
	intensity at 50\,K.
	\textsf{\textbf{b}}. This fine structure arises from transitions between a
	doublet excited state and a doublet ground state that both have zero-field
	splittings. Population is exchanged between the two excited state branches,
	but at low temperature the upward rate $\mathit{\Gamma}_\uparrow$ is slow relative
	to the state lifetime.  The downward exchange rate $\mathit{\Gamma}_\downarrow$
	dominates the relaxation within the excited state, which accounts for the
	loss of emission intensity for the high energy pair of lines.
	\textsf{\textbf{c}}. Excitation spectra of the four lines that comprise the
	SiV$^-$ ZPL (amplitudes normalised).  The higher energy lines are wider,
	consistent with a shorter effective lifetime due to thermalisation between
	the two excited state branches.
	\textsf{\textbf{d}}. The radiative decay lifetime of SiV increases from
	1.28\,ns at room temperature to 1.72\,ns at 4\,K, corresponding to a
	transform-limited PLE linewidth of 94\,MHz.
}
\label{fig:spectra}
\end{figure*}

At 4\,K the ZPL consists of four lines as a result of optically allowed
transitions between doublet ground and excited states
\cite{rogers2014electronic,hepp2014electronic} (\autoref{fig:spectra}a,b).
Each of these transitions was resonantly excited and fluorescence was detected
in the sideband. Scanning the laser frequency produced photoluminescence
excitation (PLE) spectra, shown for a single emitter in \autoref{fig:spectra}c.

The pair of lines at shorter wavelengths (higher energy), here labelled lines A
and B, had full-width at half maximum (FWHM) of 352 and 409\,MHz.  The pair of
lines at longer wavelengths were considerably narrower at 136\,MHz (line C) and
119\,MHz (line D).   All of these PLE spectra were recorded at minimal laser
intensities to avoid power-broadening.  In order to compare these linewidths to
their expected transform-limit, pulsed 532\,nm excitation was used to measure
the excited-state lifetime for several individual SiV$^-$ centers
(\autoref{fig:spectra}c).  A decay time of $1.72\pm0.04$\,ns was measured at
4\,K.  Our measured result for line D is therefore only 26\% over the
transform-limited linewidth of 94\,MHz.

It is important to account for the extra width of lines A and B, and further
information is provided by photoluminescence (PL) measured with a spectrometer.
At 4\,K lines C and D are much brighter than the higher energy lines
(\autoref{fig:spectra}a), which correspond to transitions from the upper branch
of the excited state (\autoref{fig:spectra}b). With increasing temperature
these lines gain relative intensity, indicating that thermal relaxation occurs
in the SiV$^-$ excited state. The downward exchange rate $\mathit{\Gamma}_\downarrow$
adds to the rate of decay to the ground state and reduces the effective
lifetime of the upper branch.  Consequently, lines A and B are broadened and
lose intensity in PL.  The upward exchange rate $\mathit{\Gamma}_\uparrow =
\mathit{\Gamma}_\downarrow \exp{\frac{\Delta E}{k_\mathrm{B} t}}$ depends on the Boltzmann
factor, making it small but still measurable at 4\,K.  This additional rate out
of the lower branch accounts for half of the exta linewidth above the
transform-limit for lines C and D.

At 4\,K, 71\% of the total ZPL flourescence is contained in line C.  Combined
with the Debye-Waller factor of 70\% \cite{collins1994annealing}, this means
that half of the total SiV$^-$ flourescence is emitted into the almost
transform-limited line C.  In addition, this transition is known to arise from
a single (axial) dipole moment \cite{rogers2014electronic, hepp2014electronic}.
These properties are ideal for coupling to narrowband cavities and waveguides,
and for single photon interface experiments including quantum cryptography.
The following discussion therefore focusses on line C.

\begin{figure}
\includegraphics[width=\columnwidth]{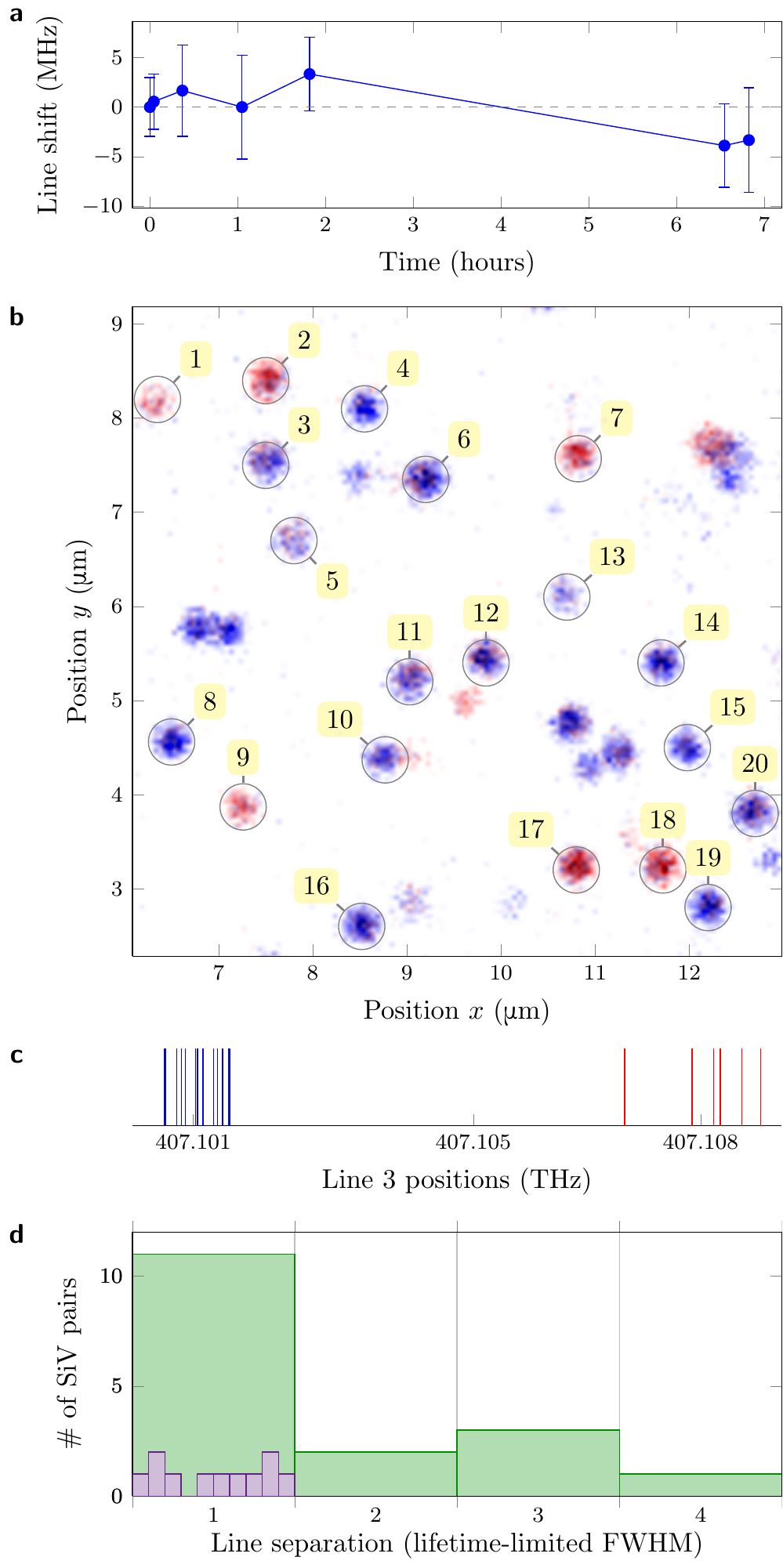}
\caption{Stability and uniformity of single SiV$^-$ centres.
	\textsf{\textbf{a}}. No measurable spectral diffusion was detected over a
	period of 7 hours.
	\textsf{\textbf{b}}. PLE spectra were measured for each of the 20
	resolvable single SiV$^-$ centres in this confocal image (numbered). The
	image is a composite of two scans performed at orthogonal excitation
	polarisations, and so the apparent colour of the sites indicates its
	orientation in the crystal lattice.
	\textsf{\textbf{c}}. Position of line C for each of the 20 SiV$^-$ centers.
	\textsf{\textbf{d}}. Histogram of the shift between adjacent sites in
	\textsf{\textbf{c}}. The primary bins are 94\,MHz wide corresponding to the
	transform-limited linewidth, and the composition of the first bin is
	illustrated in 10 sub-bins.  Eleven SiV$^-$ pairs had separations less than
	94\,MHz, and four pairs had separations less than 30\% of the
	lifetime-limited linewidth.
}
\label{fig:uniform}
\end{figure}

Over a period of 7 hours the line position was recorded for a single SiV$^-$
center (\autoref{fig:uniform}a).  The variation in line position was found to
be $\pm4$\,MHz, which is within the 95\% confidence interval for this parameter
in a Lorentzian fit.  The fact that we observe no spectral diffusion highlights
the ability of SiV$^-$ to produce indistinguishable photons over long periods
of time.  No blinking was observed for any of the SiV$^-$ sites measured in
this study.

PLE spectra were recorded for all 20 clearly-resolvable SiV$^-$ sites in an
arbitrary scan region, shown in \autoref{fig:uniform}b.  On a $\{001\}$ surface
the projections of $\langle111\rangle$-aligned SiV$^-$ centres form two
orthogonal sets \cite{rogers2014electronic}.  Scanning with two orthogonal
laser polarisations (encoded in color) revealed the set to which each site in
the region belonged \cite{rogers2014electronic}.  The line position for each
site is illustrated in \autoref{fig:uniform}c.  Within each orientation set the
sites are closely spaced, although the distinct orientations are separated by
about 5\,GHz.  \autoref{fig:uniform}d shows a histogram of these shifts between
consecutive sites.  Out of the 20 centers, 11 pairs of SiV$^-$ have separations
less than one transform-limited FWHM.  This means that for a randomly chosen
site there is more than 50\% probability of finding a second site in this scan
region whose line is displaced by less than one FWHM.  The two
closest sites had lines separated by only 6\,MHz (within the confidence interval of the fit),
meaning a spectral overlap of at least 91\%.  Notably, this spectral overlap
was achieved without external tuning of the spectral position.

To explain the observed homogeneity, we reconsider the energy level scheme in
\autoref{fig:spectra}b.  The ground and excited states have E symmetry
\cite{goss1996twelve-line, rogers2014electronic, hepp2014electronic}, and are
split due to spin-orbit interaction \cite{rogers2014electronic}.  In general,
strain and electric fields can perturb these states to result in line-shifts and
increased splittings. Electric fields may be produced by nearby charged
impurities and therefore vary across small spatial scales.  The precise
correspondence between orientation and line position for each SiV$^-$ measured
here suggests that strain, which can be nearly uniform over a
$7\times7$\,\textmu m region, is more influential than electric fields.

The large spin-orbit splittings of 46.68\,GHz (258.1\,GHz) in the ground
(excited) state help to make SiV$^-$ unresponsive to small transverse strains.
This occurs because the effect of such strain is a small perturbation
until strain splitting increases to about the magnitude of spin-orbit.  We
observed this effect in the ground-state splitting measured between lines C and
D, which varied much less ($\pm1\,$GHz) than line position across the 20 sites
and showed no correlation with orientation.  This implies that the observed
line shift results from axial strain.  The inversion symmetry of SiV$^-$
\cite{goss1996twelve-line,rogers2014electronic} reduces the influence of small axial strain,
since inverting the strain direction does not change line shift.  Our
observations indicate that this shielding has a lower threshold than provided
by spin-orbit for transverse strain.  Despite the presence of residual strain
in this sample region we were still able to find identical emitters.


\begin{figure}
\includegraphics[width=8cm]{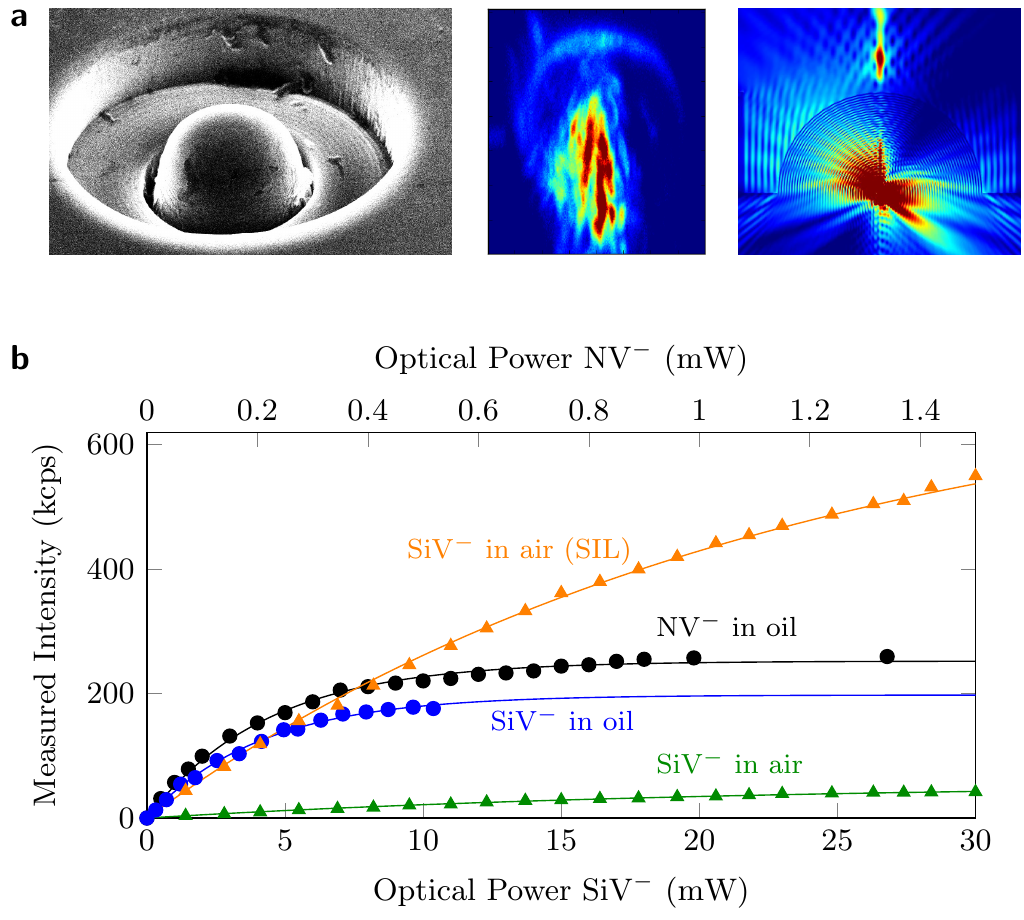}
\caption{SiV$^-$ incorporation into a SIL.
	\textsf{\textbf{a}}. Scanning electron microscope image of a single SIL
	fabricated in the diamond surface.  The vertical cross-section confocal image (center)
	shows an intensity profile in the SIL with enhanced fluorescence detection
	for SiV$^-$ centres near the focus.  This in qualitative agreement with the
	simulated emission intensity (right) for a single dipole located at the centre of
	the solid immersion lens (calculated with finite-difference time-domain
	numerical package).
	\textsf{\textbf{b}}.
	Using an air objective (NA=0.95), the saturation count-rate $I_\mathrm{sat}$ for
	SiV$^-$ under a SIL (730\,kc/s) is enhanced by more than a factor of 10
	compared to an equivalent single SiV$^-$ under a planar surface nearby
	($I_\mathrm{sat}=56$\,kc/s).
	To more easily compare SiV$^-$ with other emitters a high performance
	(NA=1.35) oil-immersion objective was used at room temperature. SiV$^-$
	sites show $I_\mathrm{sat}=200$\,kc/s which is comparable to single NV$^-$ centers
	($I_\mathrm{sat}=252$\,kc/s).
}
\label{fig:sil}
\end{figure}

An advantage to solid-state emitters over trapped ions is their reliable
addressing which allows direct incorporation into photonic and plasmonic
devices. SiV$^-$ ensembles have been coupled into cavities
\cite{lee2012coupling}, and here we incorporate single SiV$^-$ centers into
solid-immersion lens (SIL) to demonstrate their readiness for applications
(\autoref{fig:sil}a).  SILs allow increased collection of flourescence
from high refractive index materials \cite{marseglia2011nanofabricated}, which
is particularly useful during cryogenic experiments.  Well coupled SiV$^-$
centers were identified by scanning the depth profile with a confocal
microscope. The SIL enhanced flourescence collection by about a factor of ten,
giving saturation flourescence up to $I_\mathrm{sat}=730$\,kilocounts/sec (kc/s).
Using an oil-immersion objective on a flat surface, SiV$^-$ produced
$I_\mathrm{sat}=200$\,kc/s which is comparable to a single NV$^-$ centre or bright
molecule under the same conditions (\autoref{fig:sil}b).

The presence of poorly understood metastable states \cite{neu2012electronic}
prevents a deduction of absolute quantum yield for SiV$^-$ from saturation
flourescence, although it appears to be less than the quantum yield for NV$^-$
which has a flourescence lifetime of 13\,ns \cite{faraon2012coupling}.  We
found the SiV$^-$ decay lifetime to be shorter ($1.28\pm0.06$\,ns) at room
temperature than at 4\,K (\autoref{fig:spectra}d),  This is consistent with a
thermally activated non-radiative decay path \cite{feng1993characteristics,
collins1994annealing, rogers2010how}, and indicates an improvement of the
SiV$^-$ quantum yield at low temperature.  Our results show that SiV$^-$ can
provide indistinguishable photons at a collectable rate on the order of
hundreds of kc/s.


In summary, we have demonstrated a uniform single photon source in the solid
state without requiring external tuning of the optical properties.  We observe
nearly transform-limited linewidths, without spectral diffusion, which would
allow high spectral overlap between single photons emitted from distinct
sources.  The production of multiple, independent single photon emitters with
identical properties is essential to the scalability of a number of schemes
that utilise entangled photons, including quantum computing with linear optics,
and is expected to form a fundamental resource in quantum optics technologies.
The SiV$^-$ center is therefore promising for such applications.

\section*{Methods}

\begin{footnotesize}
\subsection*{Sample preparation}

$\langle001\rangle -$oriented plates cut from a low-strain, type-IIa, HPHT
crystal (Sumitomo Electric Industries, Ltd.) were used as a substrate for
microwave plasma chemical-vapour-deposition (MPCVD).  High-purity H$_2$ and
CH$_4$ source gas, specified to 99.999\% $°{12}$C isotopic enrichment
(Cambridge Isotope Laboratories CLM-392) was used to produce the plasma.  The
residual N$_2$ concentration was less than 0.1 ppb for H$_2$ and less than 1
ppb for CH$_4$. The total gas pressure, microwave power, methane concentration
ratio (CH$_4$/ H$_2$), growth duration and substrate temperature employed were
120\,Torr, 1.4\,kW, 4\,\%, 24\,h, and 950 -- 1000\,$^{\circ}$C, respectively.
The homoepitaxial layer thickness estimated from the weight difference between
initial substrate and after the growth was $\sim$ 60\,\textmu m giving a growth
rate of 2.4\,\textmu m/h.

Cathodoluminescence spectra (15\,kV acceleration voltage, $2\times10^{-7}$A
incident beam current) taken at room temperature in a wavelength range from 200
-- 800\,nm provided information on the crystalline quality and optically active
impurities. Emission from free-exciton recombination was observed from most of
the growth surface, in addition to a weak signal at 738 nm which is assigned to
SiV$^-$. The reaction vessel of MPCVD contains mainly stainless steel and
molybdenum. Quartz glass (SiO$_2$) was used for the windows of the vessel. When
homoepitaxial film was grown under low microwave power, no SiV$^-$ fluorescence
was observed. Increasing the microwave input power caused the plasma to extend
and etch material from the quartz, picking up silicon.  In this study, a 6H-SiC
single-crystal plate was also used as a Si source to allow increased silicon
doping of grown diamond.  The SiC plate was inserted between the diamond
substrate and a molybdenum sample holder.  It was possible to incorporate
SiV$^-$ during growth at concentrations below 0.2/\,\textmu m$^3$ ($\sim
1\times10^{-3}$ppb). Incorporation of silicon occurred relatively uniformly
over the whole lateral direction.

\subsection*{Optical measurements}

The sample was mounted in a continuous flow helium cryostat capable of cooling
to 4\,K.  Single SiV$^-$ centers were imaged using a home-built confocal
microscope.  The excitation laser beam (532\,nm) was focussed onto the diamond
surface through a 0.95 NA microscope objective (or 1.35 NA oil-immersion
objective for room temperature measurements).  The objective was scanned to
produce confocal images of sample regions.  Fluorescence was collected by the
same objective, filtered with a 725--775\,nm band-pass filter, and focused
through a 25\,\textmu m pinhole before detection on an avalanche photodiode (APD).
After the pinhole the flourescence could also be sent to a spectrometer  to
acquire spectra from single sites.  Using a high resolution grating (1596
grooves/mm) allowed the four component fine-structure to be resolved.

Photoluminescent exctiation (PLE) measurements were performed on the same
setup, but using a Titanium:Sapphire laser with 50\,kHz linewidth for
excitation.  This laser could scan across the entire ZPL, and detection was
performed on the sideband by switching the filter to a 750--810\,nm band-pass.

Decay lifetime of the excited state was measured by changing the excitation
laser to a Ti:Sapph pumped optical-parametric-oscillator (OPO), producing
50\,fs pulses at 80\,MHz repetition rate.  This OPO was set to 532\,nm in order
to excite SiV$^-$ off-resonantly.  The APD signal was analysed using a
PicoQuant TimeHarp counting card, with a resolution of a 25\,ps.

\subsection*{SIL fabrication}

SILs were produced in a region of the sample known to contain a medium
density of SiV$^-$ centres at a depth of 2.5-6\,\textmu m below the surface.  The
site density was high enough to assure a reasonable probability of coupling a
SIL to a colour centre.  The fabrication process was performed with a Helios
Nanolab $600$ Focus Ion Beam (FIB) lithography system.  Each SIL had a radius
of 7\,\textmu m and a depth of 7\,\textmu m.  The SIL images were taken with a Helios
Nanolab 600 Focussed Ion Beam lithography system.

\end{footnotesize}

\section*{References}
\bibliography{SiV_uniform}

\begin{thebibliography}{30}%
\makeatletter
\providecommand \@ifxundefined [1]{%
 \@ifx{#1\undefined}
}%
\providecommand \@ifnum [1]{%
 \ifnum #1\expandafter \@firstoftwo
 \else \expandafter \@secondoftwo
 \fi
}%
\providecommand \@ifx [1]{%
 \ifx #1\expandafter \@firstoftwo
 \else \expandafter \@secondoftwo
 \fi
}%
\providecommand \natexlab [1]{#1}%
\providecommand \enquote  [1]{``#1''}%
\providecommand \bibnamefont  [1]{#1}%
\providecommand \bibfnamefont [1]{#1}%
\providecommand \citenamefont [1]{#1}%
\providecommand \href@noop [0]{\@secondoftwo}%
\providecommand \href [0]{\begingroup \@sanitize@url \@href}%
\providecommand \@href[1]{\@@startlink{#1}\@@href}%
\providecommand \@@href[1]{\endgroup#1\@@endlink}%
\providecommand \@sanitize@url [0]{\catcode `\\12\catcode `\$12\catcode
  `\&12\catcode `\#12\catcode `\^12\catcode `\_12\catcode `\%12\relax}%
\providecommand \@@startlink[1]{}%
\providecommand \@@endlink[0]{}%
\providecommand \url  [0]{\begingroup\@sanitize@url \@url }%
\providecommand \@url [1]{\endgroup\@href {#1}{\urlprefix }}%
\providecommand \urlprefix  [0]{URL }%
\providecommand \Eprint [0]{\href }%
\providecommand \doibase [0]{http://dx.doi.org/}%
\providecommand \selectlanguage [0]{\@gobble}%
\providecommand \bibinfo  [0]{\@secondoftwo}%
\providecommand \bibfield  [0]{\@secondoftwo}%
\providecommand \translation [1]{[#1]}%
\providecommand \BibitemOpen [0]{}%
\providecommand \bibitemStop [0]{}%
\providecommand \bibitemNoStop [0]{.\EOS\space}%
\providecommand \EOS [0]{\spacefactor3000\relax}%
\providecommand \BibitemShut  [1]{\csname bibitem#1\endcsname}%
\let\auto@bib@innerbib\@empty
\bibitem [{\citenamefont {Knill}\ \emph {et~al.}(2001)\citenamefont {Knill},
  \citenamefont {Laflamme},\ and\ \citenamefont {Milburn}}]{knill2001scheme}%
  \BibitemOpen
  \bibfield  {author} {\bibinfo {author} {\bibfnamefont {E.}~\bibnamefont
  {Knill}}, \bibinfo {author} {\bibfnamefont {R.}~\bibnamefont {Laflamme}}, \
  and\ \bibinfo {author} {\bibfnamefont {G.~J.}\ \bibnamefont {Milburn}},\
  }\href {\doibase 10.1038/35051009} {\bibfield  {journal} {\bibinfo  {journal}
  {Nature}\ }\textbf {\bibinfo {volume} {409}},\ \bibinfo {pages} {46}
  (\bibinfo {year} {2001})}\BibitemShut {NoStop}%
\bibitem [{\citenamefont {Yuan}\ \emph {et~al.}(2008)\citenamefont {Yuan},
  \citenamefont {Chen}, \citenamefont {Zhao}, \citenamefont {Chen},
  \citenamefont {Schmiedmayer},\ and\ \citenamefont
  {Pan}}]{yuan2008experimental}%
  \BibitemOpen
  \bibfield  {author} {\bibinfo {author} {\bibfnamefont {Z.-S.}\ \bibnamefont
  {Yuan}}, \bibinfo {author} {\bibfnamefont {Y.-A.}\ \bibnamefont {Chen}},
  \bibinfo {author} {\bibfnamefont {B.}~\bibnamefont {Zhao}}, \bibinfo {author}
  {\bibfnamefont {S.}~\bibnamefont {Chen}}, \bibinfo {author} {\bibfnamefont
  {J.}~\bibnamefont {Schmiedmayer}}, \ and\ \bibinfo {author} {\bibfnamefont
  {J.-W.}\ \bibnamefont {Pan}},\ }\href {\doibase 10.1038/nature07241}
  {\bibfield  {journal} {\bibinfo  {journal} {Nature}\ }\textbf {\bibinfo
  {volume} {454}},\ \bibinfo {pages} {1098} (\bibinfo {year}
  {2008})}\BibitemShut {NoStop}%
\bibitem [{\citenamefont {Bernien}\ \emph {et~al.}(2013)\citenamefont
  {Bernien}, \citenamefont {Hensen}, \citenamefont {Pfaff}, \citenamefont
  {Koolstra}, \citenamefont {Blok}, \citenamefont {Robledo}, \citenamefont
  {Taminiau}, \citenamefont {Markham}, \citenamefont {Twitchen}, \citenamefont
  {Childress},\ and\ \citenamefont {Hanson}}]{bernien2013heralded}%
  \BibitemOpen
  \bibfield  {author} {\bibinfo {author} {\bibfnamefont {H.}~\bibnamefont
  {Bernien}}, \bibinfo {author} {\bibfnamefont {B.}~\bibnamefont {Hensen}},
  \bibinfo {author} {\bibfnamefont {W.}~\bibnamefont {Pfaff}}, \bibinfo
  {author} {\bibfnamefont {G.}~\bibnamefont {Koolstra}}, \bibinfo {author}
  {\bibfnamefont {M.~S.}\ \bibnamefont {Blok}}, \bibinfo {author}
  {\bibfnamefont {L.}~\bibnamefont {Robledo}}, \bibinfo {author} {\bibfnamefont
  {T.~H.}\ \bibnamefont {Taminiau}}, \bibinfo {author} {\bibfnamefont
  {M.}~\bibnamefont {Markham}}, \bibinfo {author} {\bibfnamefont {D.~J.}\
  \bibnamefont {Twitchen}}, \bibinfo {author} {\bibfnamefont {L.}~\bibnamefont
  {Childress}}, \ and\ \bibinfo {author} {\bibfnamefont {R.}~\bibnamefont
  {Hanson}},\ }\href {\doibase 10.1038/nature12016} {\bibfield  {journal}
  {\bibinfo  {journal} {Nature}\ }\textbf {\bibinfo {volume} {497}},\ \bibinfo
  {pages} {86} (\bibinfo {year} {2013})}\BibitemShut {NoStop}%
\bibitem [{\citenamefont {Patel}\ \emph {et~al.}(2010)\citenamefont {Patel},
  \citenamefont {Bennett}, \citenamefont {Farrer}, \citenamefont {Nicoll},
  \citenamefont {Ritchie},\ and\ \citenamefont
  {Shields}}]{patel2010two-photon}%
  \BibitemOpen
  \bibfield  {author} {\bibinfo {author} {\bibfnamefont {R.~B.}\ \bibnamefont
  {Patel}}, \bibinfo {author} {\bibfnamefont {A.~J.}\ \bibnamefont {Bennett}},
  \bibinfo {author} {\bibfnamefont {I.}~\bibnamefont {Farrer}}, \bibinfo
  {author} {\bibfnamefont {C.~A.}\ \bibnamefont {Nicoll}}, \bibinfo {author}
  {\bibfnamefont {D.~A.}\ \bibnamefont {Ritchie}}, \ and\ \bibinfo {author}
  {\bibfnamefont {A.~J.}\ \bibnamefont {Shields}},\ }\href {\doibase
  10.1038/nphoton.2010.161} {\bibfield  {journal} {\bibinfo  {journal} {Nature
  Photonics}\ }\textbf {\bibinfo {volume} {4}},\ \bibinfo {pages} {632}
  (\bibinfo {year} {2010})}\BibitemShut {NoStop}%
\bibitem [{\citenamefont {Bernien}\ \emph {et~al.}(2012)\citenamefont
  {Bernien}, \citenamefont {Childress}, \citenamefont {Robledo}, \citenamefont
  {Markham}, \citenamefont {Twitchen},\ and\ \citenamefont
  {Hanson}}]{bernien2012two-photon}%
  \BibitemOpen
  \bibfield  {author} {\bibinfo {author} {\bibfnamefont {H.}~\bibnamefont
  {Bernien}}, \bibinfo {author} {\bibfnamefont {L.}~\bibnamefont {Childress}},
  \bibinfo {author} {\bibfnamefont {L.}~\bibnamefont {Robledo}}, \bibinfo
  {author} {\bibfnamefont {M.}~\bibnamefont {Markham}}, \bibinfo {author}
  {\bibfnamefont {D.}~\bibnamefont {Twitchen}}, \ and\ \bibinfo {author}
  {\bibfnamefont {R.}~\bibnamefont {Hanson}},\ }\href {\doibase
  10.1103/PhysRevLett.108.043604} {\bibfield  {journal} {\bibinfo  {journal}
  {Physical Review Letters}\ }\textbf {\bibinfo {volume} {108}},\ \bibinfo
  {pages} {043604} (\bibinfo {year} {2012})}\BibitemShut {NoStop}%
\bibitem [{\citenamefont {Sipahigil}\ \emph {et~al.}(2012)\citenamefont
  {Sipahigil}, \citenamefont {Goldman}, \citenamefont {Togan}, \citenamefont
  {Chu}, \citenamefont {Markham}, \citenamefont {Twitchen}, \citenamefont
  {Zibrov}, \citenamefont {Kubanek},\ and\ \citenamefont
  {Lukin}}]{sipahigil2012quantum}%
  \BibitemOpen
  \bibfield  {author} {\bibinfo {author} {\bibfnamefont {A.}~\bibnamefont
  {Sipahigil}}, \bibinfo {author} {\bibfnamefont {M.~L.}\ \bibnamefont
  {Goldman}}, \bibinfo {author} {\bibfnamefont {E.}~\bibnamefont {Togan}},
  \bibinfo {author} {\bibfnamefont {Y.}~\bibnamefont {Chu}}, \bibinfo {author}
  {\bibfnamefont {M.}~\bibnamefont {Markham}}, \bibinfo {author} {\bibfnamefont
  {D.~J.}\ \bibnamefont {Twitchen}}, \bibinfo {author} {\bibfnamefont {A.~S.}\
  \bibnamefont {Zibrov}}, \bibinfo {author} {\bibfnamefont {A.}~\bibnamefont
  {Kubanek}}, \ and\ \bibinfo {author} {\bibfnamefont {M.~D.}\ \bibnamefont
  {Lukin}},\ }\href {\doibase 10.1103/PhysRevLett.108.143601} {\bibfield
  {journal} {\bibinfo  {journal} {Physical Review Letters}\ }\textbf {\bibinfo
  {volume} {108}},\ \bibinfo {pages} {143601} (\bibinfo {year}
  {2012})}\BibitemShut {NoStop}%
\bibitem [{\citenamefont {Santori}\ \emph {et~al.}(2002)\citenamefont
  {Santori}, \citenamefont {Fattal}, \citenamefont {Vu{\v c}kovi{\'c}},
  \citenamefont {Solomon},\ and\ \citenamefont
  {Yamamoto}}]{santori2002indistinguishable}%
  \BibitemOpen
  \bibfield  {author} {\bibinfo {author} {\bibfnamefont {C.}~\bibnamefont
  {Santori}}, \bibinfo {author} {\bibfnamefont {D.}~\bibnamefont {Fattal}},
  \bibinfo {author} {\bibfnamefont {J.}~\bibnamefont {Vu{\v c}kovi{\'c}}},
  \bibinfo {author} {\bibfnamefont {G.~S.}\ \bibnamefont {Solomon}}, \ and\
  \bibinfo {author} {\bibfnamefont {Y.}~\bibnamefont {Yamamoto}},\ }\href
  {\doibase 10.1038/nature01086} {\bibfield  {journal} {\bibinfo  {journal}
  {Nature}\ }\textbf {\bibinfo {volume} {419}},\ \bibinfo {pages} {594}
  (\bibinfo {year} {2002})}\BibitemShut {NoStop}%
\bibitem [{\citenamefont {Strekalov}\ \emph {et~al.}(1995)\citenamefont
  {Strekalov}, \citenamefont {Sergienko}, \citenamefont {Klyshko},\ and\
  \citenamefont {Shih}}]{strekalov1995observation}%
  \BibitemOpen
  \bibfield  {author} {\bibinfo {author} {\bibfnamefont {D.~V.}\ \bibnamefont
  {Strekalov}}, \bibinfo {author} {\bibfnamefont {A.~V.}\ \bibnamefont
  {Sergienko}}, \bibinfo {author} {\bibfnamefont {D.~N.}\ \bibnamefont
  {Klyshko}}, \ and\ \bibinfo {author} {\bibfnamefont {Y.~H.}\ \bibnamefont
  {Shih}},\ }\href {\doibase 10.1103/PhysRevLett.74.3600} {\bibfield  {journal}
  {\bibinfo  {journal} {Physical Review Letters}\ }\textbf {\bibinfo {volume}
  {74}},\ \bibinfo {pages} {3600} (\bibinfo {year} {1995})}\BibitemShut
  {NoStop}%
\bibitem [{\citenamefont {Beugnon}\ \emph {et~al.}(2006)\citenamefont
  {Beugnon}, \citenamefont {Jones}, \citenamefont {Dingjan}, \citenamefont
  {Darqui{\'e}}, \citenamefont {Messin}, \citenamefont {Browaeys},\ and\
  \citenamefont {Grangier}}]{beugnon2006quantum}%
  \BibitemOpen
  \bibfield  {author} {\bibinfo {author} {\bibfnamefont {J.}~\bibnamefont
  {Beugnon}}, \bibinfo {author} {\bibfnamefont {M.~P.~A.}\ \bibnamefont
  {Jones}}, \bibinfo {author} {\bibfnamefont {J.}~\bibnamefont {Dingjan}},
  \bibinfo {author} {\bibfnamefont {B.}~\bibnamefont {Darqui{\'e}}}, \bibinfo
  {author} {\bibfnamefont {G.}~\bibnamefont {Messin}}, \bibinfo {author}
  {\bibfnamefont {A.}~\bibnamefont {Browaeys}}, \ and\ \bibinfo {author}
  {\bibfnamefont {P.}~\bibnamefont {Grangier}},\ }\href {\doibase
  10.1038/nature04628} {\bibfield  {journal} {\bibinfo  {journal} {Nature}\
  }\textbf {\bibinfo {volume} {440}},\ \bibinfo {pages} {779} (\bibinfo {year}
  {2006})}\BibitemShut {NoStop}%
\bibitem [{\citenamefont {Shields}(2007)}]{shields2007semiconductor}%
  \BibitemOpen
  \bibfield  {author} {\bibinfo {author} {\bibfnamefont {A.~J.}\ \bibnamefont
  {Shields}},\ }\href {\doibase 10.1038/nphoton.2007.46} {\bibfield  {journal}
  {\bibinfo  {journal} {Nature Photonics}\ }\textbf {\bibinfo {volume} {1}},\
  \bibinfo {pages} {215} (\bibinfo {year} {2007})}\BibitemShut {NoStop}%
\bibitem [{\citenamefont {Kiraz}\ \emph {et~al.}(2005)\citenamefont {Kiraz},
  \citenamefont {Ehrl}, \citenamefont {Hellerer}, \citenamefont
  {M{\"u}stecapl{\i}o{\u g}lu}, \citenamefont {Br{\"a}uchle},\ and\
  \citenamefont {Zumbusch}}]{kiraz2005indistinguishable}%
  \BibitemOpen
  \bibfield  {author} {\bibinfo {author} {\bibfnamefont {A.}~\bibnamefont
  {Kiraz}}, \bibinfo {author} {\bibfnamefont {M.}~\bibnamefont {Ehrl}},
  \bibinfo {author} {\bibfnamefont {T.}~\bibnamefont {Hellerer}}, \bibinfo
  {author} {\bibfnamefont {{\"O}.~E.}\ \bibnamefont {M{\"u}stecapl{\i}o{\u
  g}lu}}, \bibinfo {author} {\bibfnamefont {C.}~\bibnamefont {Br{\"a}uchle}}, \
  and\ \bibinfo {author} {\bibfnamefont {A.}~\bibnamefont {Zumbusch}},\ }\href
  {\doibase 10.1103/PhysRevLett.94.223602} {\bibfield  {journal} {\bibinfo
  {journal} {Physical Review Letters}\ }\textbf {\bibinfo {volume} {94}},\
  \bibinfo {pages} {223602} (\bibinfo {year} {2005})}\BibitemShut {NoStop}%
\bibitem [{\citenamefont {Faraon}\ \emph {et~al.}(2012)\citenamefont {Faraon},
  \citenamefont {Santori}, \citenamefont {Huang}, \citenamefont {Acosta},\ and\
  \citenamefont {Beausoleil}}]{faraon2012coupling}%
  \BibitemOpen
  \bibfield  {author} {\bibinfo {author} {\bibfnamefont {A.}~\bibnamefont
  {Faraon}}, \bibinfo {author} {\bibfnamefont {C.}~\bibnamefont {Santori}},
  \bibinfo {author} {\bibfnamefont {Z.}~\bibnamefont {Huang}}, \bibinfo
  {author} {\bibfnamefont {V.~M.}\ \bibnamefont {Acosta}}, \ and\ \bibinfo
  {author} {\bibfnamefont {R.~G.}\ \bibnamefont {Beausoleil}},\ }\href
  {\doibase 10.1103/PhysRevLett.109.033604} {\bibfield  {journal} {\bibinfo
  {journal} {Physical Review Letters}\ }\textbf {\bibinfo {volume} {109}},\
  \bibinfo {pages} {033604} (\bibinfo {year} {2012})}\BibitemShut {NoStop}%
\bibitem [{\citenamefont {Barclay}\ \emph {et~al.}(2011)\citenamefont
  {Barclay}, \citenamefont {Fu}, \citenamefont {Santori}, \citenamefont
  {Faraon},\ and\ \citenamefont {Beausoleil}}]{barclay2011hybrid}%
  \BibitemOpen
  \bibfield  {author} {\bibinfo {author} {\bibfnamefont {P.~E.}\ \bibnamefont
  {Barclay}}, \bibinfo {author} {\bibfnamefont {K.-M.~C.}\ \bibnamefont {Fu}},
  \bibinfo {author} {\bibfnamefont {C.}~\bibnamefont {Santori}}, \bibinfo
  {author} {\bibfnamefont {A.}~\bibnamefont {Faraon}}, \ and\ \bibinfo {author}
  {\bibfnamefont {R.~G.}\ \bibnamefont {Beausoleil}},\ }\href {\doibase
  10.1103/PhysRevX.1.011007} {\bibfield  {journal} {\bibinfo  {journal}
  {Physical Review X}\ }\textbf {\bibinfo {volume} {1}},\ \bibinfo {pages}
  {011007} (\bibinfo {year} {2011})}\BibitemShut {NoStop}%
\bibitem [{\citenamefont {Collins}\ \emph {et~al.}(1990)\citenamefont
  {Collins}, \citenamefont {Kamo},\ and\ \citenamefont
  {Sato}}]{collins1990spectroscopic}%
  \BibitemOpen
  \bibfield  {author} {\bibinfo {author} {\bibfnamefont {A.~T.}\ \bibnamefont
  {Collins}}, \bibinfo {author} {\bibfnamefont {M.}~\bibnamefont {Kamo}}, \
  and\ \bibinfo {author} {\bibfnamefont {Y.}~\bibnamefont {Sato}},\ }\href
  {\doibase 10.1557/JMR.1990.2507} {\bibfield  {journal} {\bibinfo  {journal}
  {Journal of Materials Research}\ }\textbf {\bibinfo {volume} {5}},\ \bibinfo
  {pages} {2507} (\bibinfo {year} {1990})}\BibitemShut {NoStop}%
\bibitem [{\citenamefont {Goss}\ \emph {et~al.}(1996)\citenamefont {Goss},
  \citenamefont {Jones}, \citenamefont {Breuer}, \citenamefont {Briddon},\ and\
  \citenamefont {{\"O}berg}}]{goss1996twelve-line}%
  \BibitemOpen
  \bibfield  {author} {\bibinfo {author} {\bibfnamefont {J.~P.}\ \bibnamefont
  {Goss}}, \bibinfo {author} {\bibfnamefont {R.}~\bibnamefont {Jones}},
  \bibinfo {author} {\bibfnamefont {S.~J.}\ \bibnamefont {Breuer}}, \bibinfo
  {author} {\bibfnamefont {P.~R.}\ \bibnamefont {Briddon}}, \ and\ \bibinfo
  {author} {\bibfnamefont {S.}~\bibnamefont {{\"O}berg}},\ }\href {\doibase
  10.1103/PhysRevLett.77.3041} {\bibfield  {journal} {\bibinfo  {journal}
  {Physical Review Letters}\ }\textbf {\bibinfo {volume} {77}},\ \bibinfo
  {pages} {3041} (\bibinfo {year} {1996})}\BibitemShut {NoStop}%
\bibitem [{\citenamefont {Neu}\ \emph {et~al.}(2013)\citenamefont {Neu},
  \citenamefont {Hepp}, \citenamefont {Hauschild}, \citenamefont {Gsell},
  \citenamefont {Fischer}, \citenamefont {Sternschulte}, \citenamefont
  {Steinm{\"u}ller-Nethl}, \citenamefont {Schreck},\ and\ \citenamefont
  {Becher}}]{neu2013low-temperature}%
  \BibitemOpen
  \bibfield  {author} {\bibinfo {author} {\bibfnamefont {E.}~\bibnamefont
  {Neu}}, \bibinfo {author} {\bibfnamefont {C.}~\bibnamefont {Hepp}}, \bibinfo
  {author} {\bibfnamefont {M.}~\bibnamefont {Hauschild}}, \bibinfo {author}
  {\bibfnamefont {S.}~\bibnamefont {Gsell}}, \bibinfo {author} {\bibfnamefont
  {M.}~\bibnamefont {Fischer}}, \bibinfo {author} {\bibfnamefont
  {H.}~\bibnamefont {Sternschulte}}, \bibinfo {author} {\bibfnamefont
  {D.}~\bibnamefont {Steinm{\"u}ller-Nethl}}, \bibinfo {author} {\bibfnamefont
  {M.}~\bibnamefont {Schreck}}, \ and\ \bibinfo {author} {\bibfnamefont
  {C.}~\bibnamefont {Becher}},\ }\href {\doibase 10.1088/1367-2630/15/4/043005}
  {\bibfield  {journal} {\bibinfo  {journal} {New Journal of Physics}\ }\textbf
  {\bibinfo {volume} {15}},\ \bibinfo {pages} {043005} (\bibinfo {year}
  {2013})}\BibitemShut {NoStop}%
\bibitem [{\citenamefont {Feng}\ and\ \citenamefont
  {Schwartz}(1993)}]{feng1993characteristics}%
  \BibitemOpen
  \bibfield  {author} {\bibinfo {author} {\bibfnamefont {T.}~\bibnamefont
  {Feng}}\ and\ \bibinfo {author} {\bibfnamefont {B.~D.}\ \bibnamefont
  {Schwartz}},\ }\href {\doibase doi:10.1063/1.353239} {\bibfield  {journal}
  {\bibinfo  {journal} {Journal of Applied Physics}\ }\textbf {\bibinfo
  {volume} {73}},\ \bibinfo {pages} {1415} (\bibinfo {year}
  {1993})}\BibitemShut {NoStop}%
\bibitem [{\citenamefont {Brown}\ and\ \citenamefont
  {Rand}(1995)}]{brown1995site}%
  \BibitemOpen
  \bibfield  {author} {\bibinfo {author} {\bibfnamefont {S.~W.}\ \bibnamefont
  {Brown}}\ and\ \bibinfo {author} {\bibfnamefont {S.~C.}\ \bibnamefont
  {Rand}},\ }\href {\doibase doi:10.1063/1.359864} {\bibfield  {journal}
  {\bibinfo  {journal} {Journal of Applied Physics}\ }\textbf {\bibinfo
  {volume} {78}},\ \bibinfo {pages} {4069} (\bibinfo {year}
  {1995})}\BibitemShut {NoStop}%
\bibitem [{\citenamefont {Clark}\ \emph {et~al.}(1995)\citenamefont {Clark},
  \citenamefont {Kanda}, \citenamefont {Kiflawi},\ and\ \citenamefont
  {Sittas}}]{clark1995silicon}%
  \BibitemOpen
  \bibfield  {author} {\bibinfo {author} {\bibfnamefont {C.~D.}\ \bibnamefont
  {Clark}}, \bibinfo {author} {\bibfnamefont {H.}~\bibnamefont {Kanda}},
  \bibinfo {author} {\bibfnamefont {I.}~\bibnamefont {Kiflawi}}, \ and\
  \bibinfo {author} {\bibfnamefont {G.}~\bibnamefont {Sittas}},\ }\href
  {\doibase 10.1103/PhysRevB.51.16681} {\bibfield  {journal} {\bibinfo
  {journal} {Physical Review B}\ }\textbf {\bibinfo {volume} {51}},\ \bibinfo
  {pages} {16681} (\bibinfo {year} {1995})}\BibitemShut {NoStop}%
\bibitem [{\citenamefont {Edmonds}\ \emph {et~al.}(2008)\citenamefont
  {Edmonds}, \citenamefont {Newton}, \citenamefont {Martineau}, \citenamefont
  {Twitchen},\ and\ \citenamefont {Williams}}]{edmonds2008electron}%
  \BibitemOpen
  \bibfield  {author} {\bibinfo {author} {\bibfnamefont {A.~M.}\ \bibnamefont
  {Edmonds}}, \bibinfo {author} {\bibfnamefont {M.~E.}\ \bibnamefont {Newton}},
  \bibinfo {author} {\bibfnamefont {P.~M.}\ \bibnamefont {Martineau}}, \bibinfo
  {author} {\bibfnamefont {D.~J.}\ \bibnamefont {Twitchen}}, \ and\ \bibinfo
  {author} {\bibfnamefont {S.~D.}\ \bibnamefont {Williams}},\ }\href {\doibase
  10.1103/PhysRevB.77.245205} {\bibfield  {journal} {\bibinfo  {journal}
  {Physical Review B}\ }\textbf {\bibinfo {volume} {77}},\ \bibinfo {pages}
  {245205} (\bibinfo {year} {2008})}\BibitemShut {NoStop}%
\bibitem [{\citenamefont {Neu}\ \emph {et~al.}(2011)\citenamefont {Neu},
  \citenamefont {Steinmetz}, \citenamefont {Riedrich-M{\"o}ller}, \citenamefont
  {Gsell}, \citenamefont {Fischer}, \citenamefont {Schreck},\ and\
  \citenamefont {Becher}}]{neu2011single}%
  \BibitemOpen
  \bibfield  {author} {\bibinfo {author} {\bibfnamefont {E.}~\bibnamefont
  {Neu}}, \bibinfo {author} {\bibfnamefont {D.}~\bibnamefont {Steinmetz}},
  \bibinfo {author} {\bibfnamefont {J.}~\bibnamefont {Riedrich-M{\"o}ller}},
  \bibinfo {author} {\bibfnamefont {S.}~\bibnamefont {Gsell}}, \bibinfo
  {author} {\bibfnamefont {M.}~\bibnamefont {Fischer}}, \bibinfo {author}
  {\bibfnamefont {M.}~\bibnamefont {Schreck}}, \ and\ \bibinfo {author}
  {\bibfnamefont {C.}~\bibnamefont {Becher}},\ }\href {\doibase
  10.1088/1367-2630/13/2/025012} {\bibfield  {journal} {\bibinfo  {journal}
  {New Journal of Physics}\ }\textbf {\bibinfo {volume} {13}},\ \bibinfo
  {pages} {025012} (\bibinfo {year} {2011})}\BibitemShut {NoStop}%
\bibitem [{\citenamefont {Turukhin}\ \emph {et~al.}(1996)\citenamefont
  {Turukhin}, \citenamefont {Liu}, \citenamefont {Gorokhovsky}, \citenamefont
  {Alfano},\ and\ \citenamefont {Phillips}}]{turukhin1996picosecond}%
  \BibitemOpen
  \bibfield  {author} {\bibinfo {author} {\bibfnamefont {A.~V.}\ \bibnamefont
  {Turukhin}}, \bibinfo {author} {\bibfnamefont {C.-H.}\ \bibnamefont {Liu}},
  \bibinfo {author} {\bibfnamefont {A.~A.}\ \bibnamefont {Gorokhovsky}},
  \bibinfo {author} {\bibfnamefont {R.~R.}\ \bibnamefont {Alfano}}, \ and\
  \bibinfo {author} {\bibfnamefont {W.}~\bibnamefont {Phillips}},\ }\href
  {\doibase 10.1103/PhysRevB.54.16448} {\bibfield  {journal} {\bibinfo
  {journal} {Physical Review B}\ }\textbf {\bibinfo {volume} {54}},\ \bibinfo
  {pages} {16448} (\bibinfo {year} {1996})}\BibitemShut {NoStop}%
\bibitem [{\citenamefont {Wang}\ \emph {et~al.}(2006)\citenamefont {Wang},
  \citenamefont {Kurtsiefer}, \citenamefont {Weinfurter},\ and\ \citenamefont
  {Burchard}}]{wang2006single}%
  \BibitemOpen
  \bibfield  {author} {\bibinfo {author} {\bibfnamefont {C.}~\bibnamefont
  {Wang}}, \bibinfo {author} {\bibfnamefont {C.}~\bibnamefont {Kurtsiefer}},
  \bibinfo {author} {\bibfnamefont {H.}~\bibnamefont {Weinfurter}}, \ and\
  \bibinfo {author} {\bibfnamefont {B.}~\bibnamefont {Burchard}},\ }\href
  {\doibase 10.1088/0953-4075/39/1/005} {\bibfield  {journal} {\bibinfo
  {journal} {Journal of Physics B: Atomic, Molecular and Optical Physics}\
  }\textbf {\bibinfo {volume} {39}},\ \bibinfo {pages} {37} (\bibinfo {year}
  {2006})}\BibitemShut {NoStop}%
\bibitem [{\citenamefont {Rogers}\ \emph {et~al.}(2014)\citenamefont {Rogers},
  \citenamefont {Jahnke}, \citenamefont {Doherty}, \citenamefont {Dietrich},
  \citenamefont {{McGuinness}}, \citenamefont {M{\"u}ller}, \citenamefont
  {Teraji}, \citenamefont {Sumiya}, \citenamefont {Isoya}, \citenamefont
  {Manson},\ and\ \citenamefont {Jelezko}}]{rogers2014electronic}%
  \BibitemOpen
  \bibfield  {author} {\bibinfo {author} {\bibfnamefont {L.~J.}\ \bibnamefont
  {Rogers}}, \bibinfo {author} {\bibfnamefont {K.~D.}\ \bibnamefont {Jahnke}},
  \bibinfo {author} {\bibfnamefont {M.~W.}\ \bibnamefont {Doherty}}, \bibinfo
  {author} {\bibfnamefont {A.}~\bibnamefont {Dietrich}}, \bibinfo {author}
  {\bibfnamefont {L.~P.}\ \bibnamefont {{McGuinness}}}, \bibinfo {author}
  {\bibfnamefont {C.}~\bibnamefont {M{\"u}ller}}, \bibinfo {author}
  {\bibfnamefont {T.}~\bibnamefont {Teraji}}, \bibinfo {author} {\bibfnamefont
  {H.}~\bibnamefont {Sumiya}}, \bibinfo {author} {\bibfnamefont
  {J.}~\bibnamefont {Isoya}}, \bibinfo {author} {\bibfnamefont {N.~B.}\
  \bibnamefont {Manson}}, \ and\ \bibinfo {author} {\bibfnamefont
  {F.}~\bibnamefont {Jelezko}},\ }\href {\doibase 10.1103/PhysRevB.89.235101}
  {\bibfield  {journal} {\bibinfo  {journal} {Physical Review B}\ }\textbf
  {\bibinfo {volume} {89}},\ \bibinfo {pages} {235101} (\bibinfo {year}
  {2014})}\BibitemShut {NoStop}%
\bibitem [{\citenamefont {Hepp}\ \emph {et~al.}(2014)\citenamefont {Hepp},
  \citenamefont {M{\"u}ller}, \citenamefont {Waselowski}, \citenamefont
  {Becker}, \citenamefont {Pingault}, \citenamefont {Sternschulte},
  \citenamefont {Steinm{\"u}ller-Nethl}, \citenamefont {Gali}, \citenamefont
  {Maze}, \citenamefont {Atat{\"u}re},\ and\ \citenamefont
  {Becher}}]{hepp2014electronic}%
  \BibitemOpen
  \bibfield  {author} {\bibinfo {author} {\bibfnamefont {C.}~\bibnamefont
  {Hepp}}, \bibinfo {author} {\bibfnamefont {T.}~\bibnamefont {M{\"u}ller}},
  \bibinfo {author} {\bibfnamefont {V.}~\bibnamefont {Waselowski}}, \bibinfo
  {author} {\bibfnamefont {J.~N.}\ \bibnamefont {Becker}}, \bibinfo {author}
  {\bibfnamefont {B.}~\bibnamefont {Pingault}}, \bibinfo {author}
  {\bibfnamefont {H.}~\bibnamefont {Sternschulte}}, \bibinfo {author}
  {\bibfnamefont {D.}~\bibnamefont {Steinm{\"u}ller-Nethl}}, \bibinfo {author}
  {\bibfnamefont {A.}~\bibnamefont {Gali}}, \bibinfo {author} {\bibfnamefont
  {J.~R.}\ \bibnamefont {Maze}}, \bibinfo {author} {\bibfnamefont
  {M.}~\bibnamefont {Atat{\"u}re}}, \ and\ \bibinfo {author} {\bibfnamefont
  {C.}~\bibnamefont {Becher}},\ }\href {\doibase
  10.1103/PhysRevLett.112.036405} {\bibfield  {journal} {\bibinfo  {journal}
  {Physical Review Letters}\ }\textbf {\bibinfo {volume} {112}},\ \bibinfo
  {pages} {036405} (\bibinfo {year} {2014})}\BibitemShut {NoStop}%
\bibitem [{\citenamefont {Collins}\ \emph {et~al.}(1994)\citenamefont
  {Collins}, \citenamefont {Allers}, \citenamefont {Wort},\ and\ \citenamefont
  {Scarsbrook}}]{collins1994annealing}%
  \BibitemOpen
  \bibfield  {author} {\bibinfo {author} {\bibfnamefont {A.~T.}\ \bibnamefont
  {Collins}}, \bibinfo {author} {\bibfnamefont {L.}~\bibnamefont {Allers}},
  \bibinfo {author} {\bibfnamefont {C.~J.}\ \bibnamefont {Wort}}, \ and\
  \bibinfo {author} {\bibfnamefont {G.~A.}\ \bibnamefont {Scarsbrook}},\ }\href
  {\doibase 10.1016/0925-9635(94)90302-6} {\bibfield  {journal} {\bibinfo
  {journal} {Diamond and Related Materials}\ }\textbf {\bibinfo {volume} {3}},\
  \bibinfo {pages} {932} (\bibinfo {year} {1994})}\BibitemShut {NoStop}%
\bibitem [{\citenamefont {Lee}\ \emph {et~al.}(2012)\citenamefont {Lee},
  \citenamefont {Aharonovich}, \citenamefont {Magyar}, \citenamefont {Rol},\
  and\ \citenamefont {Hu}}]{lee2012coupling}%
  \BibitemOpen
  \bibfield  {author} {\bibinfo {author} {\bibfnamefont {J.~C.}\ \bibnamefont
  {Lee}}, \bibinfo {author} {\bibfnamefont {I.}~\bibnamefont {Aharonovich}},
  \bibinfo {author} {\bibfnamefont {A.~P.}\ \bibnamefont {Magyar}}, \bibinfo
  {author} {\bibfnamefont {F.}~\bibnamefont {Rol}}, \ and\ \bibinfo {author}
  {\bibfnamefont {E.~L.}\ \bibnamefont {Hu}},\ }\href {\doibase
  10.1364/OE.20.008891} {\bibfield  {journal} {\bibinfo  {journal} {Optics
  Express}\ }\textbf {\bibinfo {volume} {20}},\ \bibinfo {pages} {8891}
  (\bibinfo {year} {2012})}\BibitemShut {NoStop}%
\bibitem [{\citenamefont {Marseglia}\ \emph {et~al.}(2011)\citenamefont
  {Marseglia}, \citenamefont {Hadden}, \citenamefont {Stanley-Clarke},
  \citenamefont {Harrison}, \citenamefont {Patton}, \citenamefont {Ho},
  \citenamefont {Naydenov}, \citenamefont {Jelezko}, \citenamefont {Meijer},
  \citenamefont {Dolan}, \citenamefont {Smith}, \citenamefont {Rarity},\ and\
  \citenamefont {{O{\textquoteright}Brien}}}]{marseglia2011nanofabricated}%
  \BibitemOpen
  \bibfield  {author} {\bibinfo {author} {\bibfnamefont {L.}~\bibnamefont
  {Marseglia}}, \bibinfo {author} {\bibfnamefont {J.~P.}\ \bibnamefont
  {Hadden}}, \bibinfo {author} {\bibfnamefont {A.~C.}\ \bibnamefont
  {Stanley-Clarke}}, \bibinfo {author} {\bibfnamefont {J.~P.}\ \bibnamefont
  {Harrison}}, \bibinfo {author} {\bibfnamefont {B.}~\bibnamefont {Patton}},
  \bibinfo {author} {\bibfnamefont {Y.-L.~D.}\ \bibnamefont {Ho}}, \bibinfo
  {author} {\bibfnamefont {B.}~\bibnamefont {Naydenov}}, \bibinfo {author}
  {\bibfnamefont {F.}~\bibnamefont {Jelezko}}, \bibinfo {author} {\bibfnamefont
  {J.}~\bibnamefont {Meijer}}, \bibinfo {author} {\bibfnamefont {P.~R.}\
  \bibnamefont {Dolan}}, \bibinfo {author} {\bibfnamefont {J.~M.}\ \bibnamefont
  {Smith}}, \bibinfo {author} {\bibfnamefont {J.~G.}\ \bibnamefont {Rarity}}, \
  and\ \bibinfo {author} {\bibfnamefont {J.~L.}\ \bibnamefont
  {{O{\textquoteright}Brien}}},\ }\href {\doibase doi:10.1063/1.3573870}
  {\bibfield  {journal} {\bibinfo  {journal} {Applied Physics Letters}\
  }\textbf {\bibinfo {volume} {98}},\ \bibinfo {pages} {133107} (\bibinfo
  {year} {2011})}\BibitemShut {NoStop}%
\bibitem [{\citenamefont {Neu}\ \emph {et~al.}(2012)\citenamefont {Neu},
  \citenamefont {Albrecht}, \citenamefont {Fischer}, \citenamefont {Gsell},
  \citenamefont {Schreck},\ and\ \citenamefont {Becher}}]{neu2012electronic}%
  \BibitemOpen
  \bibfield  {author} {\bibinfo {author} {\bibfnamefont {E.}~\bibnamefont
  {Neu}}, \bibinfo {author} {\bibfnamefont {R.}~\bibnamefont {Albrecht}},
  \bibinfo {author} {\bibfnamefont {M.}~\bibnamefont {Fischer}}, \bibinfo
  {author} {\bibfnamefont {S.}~\bibnamefont {Gsell}}, \bibinfo {author}
  {\bibfnamefont {M.}~\bibnamefont {Schreck}}, \ and\ \bibinfo {author}
  {\bibfnamefont {C.}~\bibnamefont {Becher}},\ }\href {\doibase
  10.1103/PhysRevB.85.245207} {\bibfield  {journal} {\bibinfo  {journal}
  {Physical Review B}\ }\textbf {\bibinfo {volume} {85}},\ \bibinfo {pages}
  {245207} (\bibinfo {year} {2012})}\BibitemShut {NoStop}%
\bibitem [{\citenamefont {Rogers}(2010)}]{rogers2010how}%
  \BibitemOpen
  \bibfield  {author} {\bibinfo {author} {\bibfnamefont {L.}~\bibnamefont
  {Rogers}},\ }\href {\doibase 10.1016/j.phpro.2010.01.221} {\bibfield
  {journal} {\bibinfo  {journal} {Physics Procedia}\ }\textbf {\bibinfo
  {volume} {3}},\ \bibinfo {pages} {1557} (\bibinfo {year} {2010})}\BibitemShut
  {NoStop}%
\end{thebibliography}%

\section*{Acknowledgements}

We acknowledge ERC, EU projects (SIQS, DIADEMS, EQUAM), DFG (FOR 1482, FOR 1493 and SFBTR 21), JST,BMBF, the
Alexander von Humboldt, and the Sino-German and Volkswagen foundations for funding. We acknowledge G. Neusser and the FIB Center UUlm for support manufacturing SILs.


\end{document}